\documentclass{ijuc}
\usepackage[english]{babel}
\usepackage{graphicx}

\newcommand {\ket} [1] {| #1 \rangle}

\newcommand {\mh}  {\hspace{.3 in}}

\newcommand {\sv}  {\smallskip}
\newcommand {\mv}  {\medskip}
\newcommand {\bv}  {\bigskip}

\newcommand {\qq} {``}

\begin{document}



\title{Diagrams of States in Quantum Information: \\ an Illustrative Tutorial}


\author{Sara Felloni$^1$\email{felloni@disco.unimib.it}
   \and Alberto Leporati$^1$
   \and Giuliano Strini$^2$}

\institute{Dipartimento di Informatica, Sistemistica e Comunicazione \\
   Universit\`a degli Studi di Milano -- Bicocca\\
Viale Sarca 336/14, 20126 Milano, Italy\\
\mv
\and
Dipartimento di Fisica\\
Universit\`a degli Studi di Milano\\
Via Celoria 16, 20133 Milano, Italy\\
}

\maketitle

\begin{abstract}


We present \textit{Diagrams of States}, a way to graphically represent and analyze how quantum information is elaborated during the execution of quantum circuits.\\
This introductory tutorial illustrates the basics, providing useful examples of quantum computations: elementary operations in single-qubit, two-qubit and three-qubit systems, immersions of gates on higher dimensional spaces, generation of single and multi-qubit states, procedures to synthesize unitary, controlled and diagonal matrices.\\
To perform the analysis of quantum processes, we directly derive diagrams of states from physical implementations of quantum circuits associated to the processes. \textit{Complete} diagrams are then rearranged into \textit{simplified} diagrams, to visualize the overall effects of computations.
Conversely, diagrams of states help to conceive new quantum algorithms, by schematically describing desired manipulations of quantum information with intuitive diagrams and then by guessing the equivalent complete diagrams, from which the corresponding quantum circuit is obtained effortlessly. Related examples and analysis of complex algorithms will be provided in future works, for whose comprehension this first tutorial offers the necessary introduction.

\end{abstract}


\bv

\section{The Graphic Representation of States: Introduction to the Method}
In this tutorial we illustrate a graphic representation of quantum information, which is new to the best of our knowledge and which we call \textit{Diagrams of States}.

Diagrams of states graphically represent and analyze how quantum information is elaborated during the execution of quantum circuits. Thus, the diagrams of states can serve as an alternative approach to analyze known quantum algorithms, as well as an auxiliary tool to conceive novel quantum computations. This introductory tutorial illustrates the basics of such a graphic representation, which can be used in addition to traditional tools such as analytical study and Feynman diagrams, as these representations are too synthetic to clearly visualize quantum information flow during computations.

The method of diagrams of states has already proven useful to study and compare some models of quantum copying machines \cite{FeSt06}. Thus we offer here a complete and detailed illustration of this novel representation.

The diagrams of states will be illustrated by means of many useful examples of quantum computations and several applications, such as elementary operations in single-qubit, two-qubit and three-qubit systems, immersions of quantum gates on spaces of higher dimensions, generation of single and multi-qubit states, procedures to synthesize general unitary, controlled and diagonal matrices.

In order to perform the analysis of quantum processes, we will directly derive the diagrams of states from the physical implementation of the quantum circuits associated to the processes. These  diagrams can easily be rearranged into new simpler diagrams, which better visualize the overall effects of the computations. We will thus call the former \textit{complete diagrams} and the latter \textit{simplified} diagrams.

\sv

This tutorial is organized as follows. Sections~\ref{sez-sd-I-1q}, \ref{sez-sd-I-2q} and \ref{sez-sd-I-3q} illustrate elementary operations performed in single-qubit, two-qubit and three-qubit systems, respectively: We present the diagrams of states for the main elementary gates in quantum computation and for immersions of these gates on systems composed of a higher number of qubits. In Section~\ref{sez-sd-I-sintucont} we present controlled gates, with control from a single qubit or a couple of qubits, and some possible procedures to synthesize this class of quantum gates. This in turn will allow one to synthesize general unitary matrices acting on a two-qubit system. Section~\ref{sez-sd-I-sintudiag} illustrates some possible procedures to synthesize general diagonal matrices acting on two-qubit and three-qubit systems and a method to synthesize general states of two and three qubits. Finally, in Section~\ref{sez-sd-I-concl} we present our conclusions.

In all the following quantum circuits and their representations by means of diagrams of states, any sequence of logic gates must be read from the left (input) to the
right (output); from top to bottom, qubits run from the least significant (\textsc{lsb}) to the most significant (\textsc{msb}).

\section{Elementary Single-qubit Operations}
\label{sez-sd-I-1q}
A single qubit constitutes a two-state quantum system. The well known following elementary operations can be performed:
\begin{itemize}
    \item the $\mbox{\sc not}$ gate;
    \item a general unitary matrix,
    $$U = \left[
\begin{array}{cc}
a & b \\
c & d
\end{array}\right],
$$
which includes the $\mbox{\sc not}$ gate as a special case.
\end{itemize}

\begin{figure}[htb]
\begin{center}
\includegraphics[width=9.8cm]{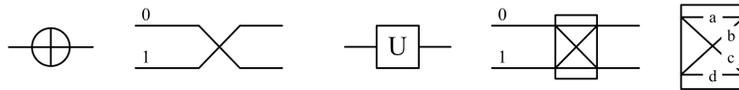}
\end{center}
\caption{Single-qubit quantum operations and the corresponding elementary diagrams of states: from left to right, quantum circuits and diagrams of states of the $\mbox{\sc not}$ gate application and of a general unitary matrix. The unitary matrix is represented by four intersecting lines, each one labeled by the corresponding entry of the matrix (see the rightmost representation).} \label{sd-I-diast1q}
\end{figure}

Figure \ref{sd-I-diast1q} illustrates the corresponding elementary representations by means of diagrams of states.

In Figure \ref{sd-I-diast1q}, the diagram of states of the $\mbox{\sc not}$ gate illustrates the exchange of the states. The unitary matrix is represented by four intersecting lines, each one labeled by the corresponding entry of the matrix; this scheme will become particularly useful in showing constructive and destructive interferences of quantum information flow in more complex diagrams.

\subsection{The Hadamard and Pauli gates}
Following the representation of a general single-qubit unitary matrix, we illustrate by means of diagrams of states other significant single-qubit elementary gates:

\begin{itemize}
    \item the Hadamard gate, $$H = \frac{1}{\sqrt2} \left[
\begin{array}{cc}
1 & 1 \\
1 & - 1
\end{array}\right];
$$
    \item the Pauli gates, $$\sigma_x = \left[
\begin{array}{cc}
0 & 1 \\
1 & 0
\end{array}\right], \mh \sigma_y = \left[
\begin{array}{cc}
0 & - i \\
i & 0
\end{array}\right], \mh \sigma_z = \left[
\begin{array}{cc}
1 & 0 \\
0 & - 1
\end{array}\right].$$
\end{itemize}

\begin{figure}[htb]
\begin{center}
\includegraphics[width=8.4cm]{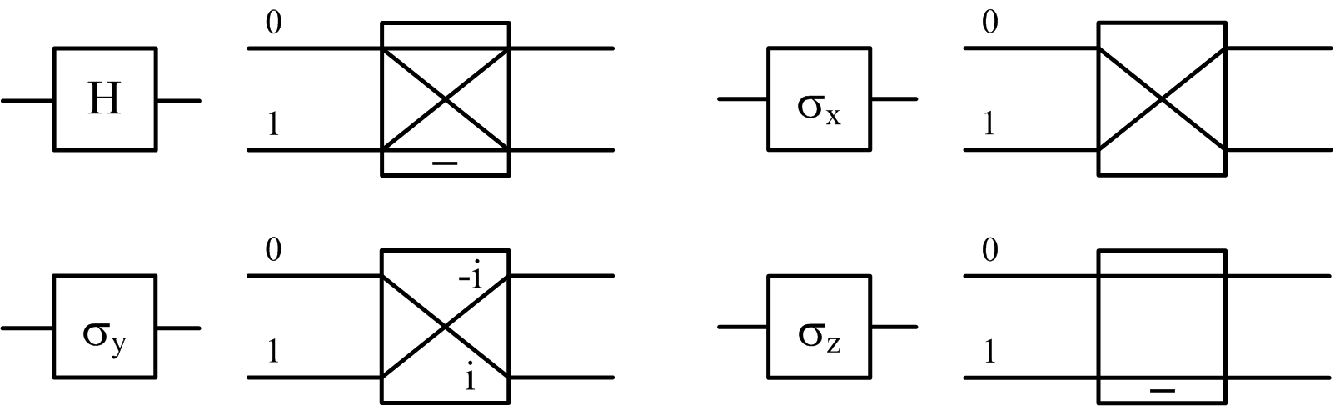}
\end{center}
  \caption{Single-qubit quantum operations and the corresponding elementary diagrams of states: from left to right, from top to bottom, the Hadamard gate and Pauli gates. For the sake of clarity, normalization coefficients are omitted in the representation of the Hadamard gate and the following notation is adopted, here and in the following: No label on the line corresponds to a \qq $+1$'' entry of the matrix, while the \qq $-$'' label corresponds to a \qq $-1$'' entry of the matrix.} \label{sd-I-diastHaPa}
\end{figure}

The diagram of states of the Hadamard gate, in Figure \ref{sd-I-diastHaPa}, shows the usual four intersecting lines in which the labels are simplified as follows, for the sake of clarity: Normalization coefficients are omitted, no label on the line corresponds to a \qq $1$'' entry of the matrix and the \qq $-$'' label corresponds to a \qq $-1$'' entry of the matrix. This simplification, adopted also for the Pauli gates, will also occur whenever useful in all the following diagrams of states.

\subsection{Phase-shift gates}
\begin{figure}[htb]
\begin{center}
\includegraphics[width=3.6cm]{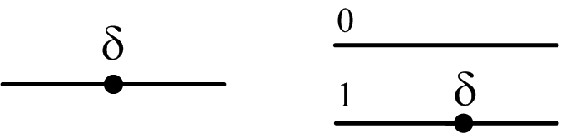}
\end{center}
  \caption{Single-qubit quantum operations and the corresponding elementary diagrams of states: the elementary phase-shift gate.}  \label{sd-I-diastph}
\end{figure}

The elementary phase-shift gate is defined by:
\begin{equation}
    \delta = \left[\begin{array}{cc}1&0\\0&e^{i\delta}\end{array}\right],
\end{equation}
to which the quantum circuit and diagram of states illustrated in Figure \ref{sd-I-diastph} can be associated.

It is important to recall that any unitary or special unitary operation on a single qubit can be constructed by using only Hadamard and phase-shift gates, since a common phase is arbitrary, as stated by the postulates of Quantum Mechanics. This is illustrated in the following sections.

\subsection{Synthesis of single-qubit special unitary matrices}
Let us now present a simple quantum circuit for the synthesis of a general special unitary matrix acting on a single-qubit system.

\begin{figure}[htb]
\begin{center}
\includegraphics[width=9.6cm]{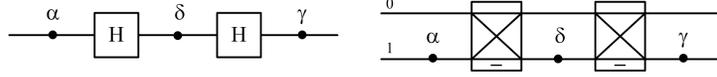}
\end{center}
  \caption{Single-qubit quantum operations and the corresponding diagrams of states: synthesis of a special unitary matrix.}
\label{sd-I-diastU1q}
\end{figure}

This synthesis can be obtained by alternate applications of three phase-shift gates and two Hadamard gates, as shown
in the quantum circuit and in the diagram of states of Figure \ref{sd-I-diastU1q}.

Applying the sequence of quantum gates, we obtain:
\begin{equation}\label{eq-sd-I-U1q}
    U = \gamma \; H \; \delta \; H \; \alpha =
    e^{i \frac{\gamma + \delta + \alpha}{2}}
    \left[
      \begin{array}{cc}
        e^{- i \frac{\gamma + \alpha}{2}} \cos \frac{\delta}{2} &
        -i e^{- i \frac{\gamma - \alpha}{2}} \sin \frac{\delta}{2} \\
        -i e^{ i \frac{\gamma - \alpha}{2}} \sin \frac{\delta}{2} &
        e^{ i \frac{\gamma + \alpha}{2}} \cos \frac{\delta}{2} \\
      \end{array}
    \right].
\end{equation}

Since the phase factor given by $e^{i \frac{\gamma + \delta +
\alpha}{2}}$ can be ignored, because a common phase is arbitrary, as stated by the postulates of Quantum Mechanics, the matrix $U$ in Eq. (\ref{eq-sd-I-U1q}) may be assumed to have determinant $=1$; hence it is a special unitary matrix, \textit{i.e.} it belongs to the algebraic group $SU(2)$, and it is general in the sense that it depends on three real parameters.

\subsection{Generation of single-qubit states}
Since a common phase factor is arbitrary, a general state $\ket\Psi$ of a single-qubit system can be generated by a quantum circuit involving a rotation of an angle $\theta$ about the $y$ axis of the Bloch sphere and by a phase-shift gate, acting on the initial state $\ket0$ (see, \textit{e.g.}, \cite{qcbook}, Ch. 3, page 111):
\begin{equation}
\ket\Psi=
\left[
\begin{array}{c}
\cos\frac{\theta}{2}\\
e^{i\delta} \sin\frac{\theta}{2}
\end{array}
\right] \equiv \left[
\begin{array}{cc}
1&0\\
0&e^{i\delta}
\end{array}
\right]  \left[
\begin{array}{cc}
\cos\frac{\theta}{2}&-\sin\frac{\theta}{2}
\\\sin\frac{\theta}{2}&\cos\frac{\theta}{2}
\end{array}
\right]  \left[
\begin{array}{l}
1\\0
\end{array}
\right].
\end{equation}

\begin{figure}[htb]
\begin{center}
\includegraphics[width=8.4cm]{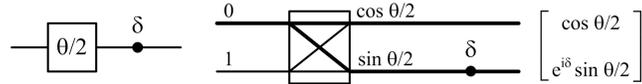}
\end{center}
  \caption{Single-qubit quantum operations and the corresponding diagrams of states: generation of a single-qubit state. Information flows on the marked lines, starting from the initial state, from left to right, while thinner lines correspond to absence of information.}
\label{sd-I-diastst1q}
\end{figure}

Figure \ref{sd-I-diastst1q} shows the quantum circuit and the corresponding diagram of states for the generation of a single-qubit state. Information flows on the marked lines, from left to right, starting from the initial state set to $\ket0$, while thinner lines correspond to absence of information.

\section{Elementary Two-qubit Operations}\label{sez-sd-I-2q}
Two qubits constitute a four-state quantum system. In the following, we illustrate elementary operations that can be performed on two qubits and the corresponding diagrams of states. First, we consider the main two-qubit gates; subsequently, we illustrate immersions of single-qubit gates on the space of two qubits; finally, we introduce controlled gates.

The generation of states of a two-qubit system will be presented in section \ref{sez-sd-I-sint2q3q}.

\subsection{Generalized \mbox{\sc cnot} gates and the \mbox{\sc swap} gate}
\begin{figure}[htb]
\begin{center}
\includegraphics[width=9cm]{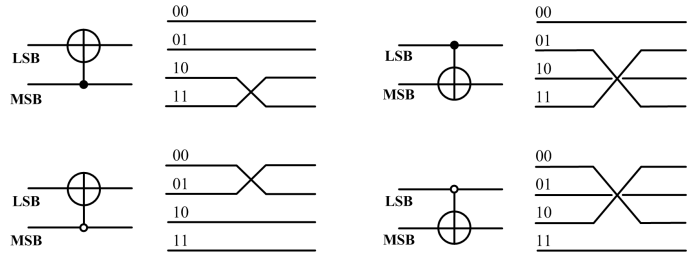}
\end{center}
  \caption{Two-qubit quantum operations and the corresponding elementary diagrams of states: generalized $\mbox{\sc cnot}$ gates. From left to right, from top to bottom, the $\mbox{\sc cnot}$, $\mbox{\sc cnot-r}$, $\overline{\mbox{\sc cnot}}$ and $\overline{\mbox{\sc cnot-r}}$ gates are shown. Here and in the following representations by means of diagrams of states, lines corresponding to switches of states do intersect, while states that do not switch correspond to overlapping (not intersecting) lines.}
\label{sd-I-diastcnot}
\end{figure}

\begin{figure}[htb]
\begin{center}
\includegraphics[width=9.4cm]{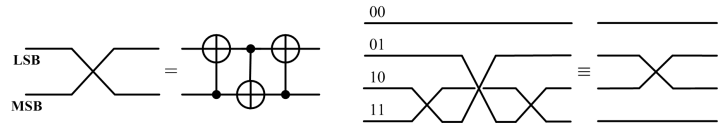}
\end{center}
  \caption{Two-qubit quantum operations and the corresponding elementary diagrams of states: representation and a possible synthesis of the $\mbox{\sc swap}$ gate.}
\label{sd-I-diastswap}
\end{figure}

We show the usual representations and elementary diagrams of states of the four possible generalized $\mbox{\sc cnot}$ gates and of the $\mbox{\sc swap}$ gate, as illustrated in Figures \ref{sd-I-diastcnot} and \ref{sd-I-diastswap}:

$$
    \mbox{\sc cnot} = \left[%
\begin{array}{cccc}
  1 & 0 & 0 & 0 \\
  0 & 1 & 0 & 0 \\
  0 & 0 & 0 & 1 \\
  0 & 0 & 1 & 0 \\
\end{array}%
\right]; \mh \overline{\textit{\mbox{\sc cnot}}} = \left[%
\begin{array}{cccc}
  0 & 1 & 0 & 0 \\
  1 & 0 & 0 & 0 \\
  0 & 0 & 1 & 0 \\
  0 & 0 & 0 & 1 \\
\end{array}%
\right];
$$
$$
\mbox{\sc cnot-r} = \left[%
\begin{array}{cccc}
  1 & 0 & 0 & 0 \\
  0 & 0 & 0 & 1 \\
  0 & 0 & 1 & 0 \\
  0 & 1 & 0 & 0 \\
\end{array}%
\right]; \mh \overline{\mbox{\sc cnot-r}} = \left[%
\begin{array}{cccc}
  0 & 0 & 1 & 0 \\
  0 & 1 & 0 & 0 \\
  1 & 0 & 0 & 0 \\
  0 & 0 & 0 & 1 \\
\end{array}%
\right];
$$
\begin{equation}
    \mbox{\sc swap} = \left[%
\begin{array}{cccc}
  1 & 0 & 0 & 0 \\
  0 & 0 & 1 & 0 \\
  0 & 1 & 0 & 0 \\
  0 & 0 & 0 & 1 \\
\end{array}%
\right].
\end{equation}

As clearly illustrated by the corresponding diagrams of states, the $\mbox{\sc cnot}$ gate switches the states that correspond to
$\textsc{msb} = 1$, that is the couple $\{ 10, 11 \}$. The $\overline{\mbox{\sc cnot}}$ gate performs the same operation for $\textsc{msb} = 0$ and thus the states $\{ 00, 01 \}$ are switched. The
$\mbox{\sc cnot-r}$ and $\overline{\mbox{\sc cnot-r}}$ gates perform similar operations with the control now set on the least significant qubit (instead of the most significant qubit): The first gate switches the couple of states $\{ 01, 11 \}$, while the second gate switches the couple of states $\{ 00, 10 \}$. As before, states that do not switch correspond to overlapping (not intersecting) lines. Finally, the \emph{\mbox{\sc swap}} gate switches the states $01$ and $10$, leaving the states $00$ and $11$ unchanged.

\subsection{Immersions of single-qubit gates on the space of two qubits}
One of the fundamental results concerning the feasibility of arbitrary quantum computations is that any unitary matrix can be decomposed into the appropriate combination of unitary matrices acting on a single qubit and $\mbox{\sc cnot}$ gates, or by means of equivalent methods, depending on the particular physical implementation of the quantum computer.

Every decomposition of a general unitary matrix into simpler matrices can be obtained by considering an appropriate collection of elementary matrices, set into the spaces constituted by a higher number of qubits. Thus we illustrate the analytic representation, usual quantum circuits and diagrams of states for immersion operations.

The diagrams of states illustrate the information flow in tensor products of matrices. Clearly, the distribution of information in the diagram is determined by the \qq weight'' of the qubit to which the operation is applied, where the qubit's \qq weight'' means the qubit's position between the most significant and the least significant ones.

In this section we illustrate the immersions of single-qubit gates on the space of two qubits. There are two possible cases, since the single-qubit gate can be applied to the most significant or to the least significant qubit.

\begin{figure}[htb]
\begin{center}
\includegraphics[width=9cm]{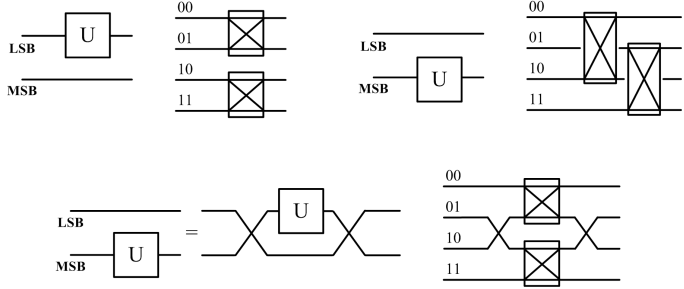}
\end{center}
  \caption{Two-qubit quantum operations and the corresponding elementary diagrams of states: immersions of single-qubit gates on the space of two qubits. From left to right, the single-qubit gate is applied to the least significant qubit and to the most significant qubit (top). The bottom representation shows an alternative diagram for the single-qubit gate applied to the most significant qubit, by applying two $\mbox{\sc swap}$ gates.} \label{sd-I-diastimm1to2q}
\end{figure}

For a general unitary matrix:
\begin{equation}
U = \left[
\begin{array}{cc}
a & b \\
c & d
\end{array}\right],
\end{equation}
we obtain the two possible immersions, illustrated in Figure \ref{sd-I-diastimm1to2q}:
\begin{itemize}
    \item unitary matrix applied to the least significant qubit,
\begin{equation}
    U_{LSB} = I \otimes U = \left[%
\begin{array}{cccc}
  U & 0  \\
  0 & U \\
\end{array}%
\right] = \left[
\begin{array}{cccc}
a&b&0&0\\
c&d&0&0\\
0&0&a&b\\
0&0&c&d
\end{array}
\right];
\end{equation}
    \item unitary matrix applied to the most significant qubit, whose analytical description can be obtained by applying two $\mbox{\sc swap}$ gates and following the former scheme, relative to the unitary matrix applied to the least significant qubit,
$$
    U_{MSB} = \left[%
\begin{array}{cccc}
  1 & 0 & 0 & 0 \\
  0 & 0 & 1 & 0 \\
  0 & 1 & 0 & 0 \\
  0 & 0 & 0 & 1 \\
\end{array}%
\right] \left[
\begin{array} {cc}
U & 0 \\
0 & U
\end{array}
\right]\left[%
\begin{array}{cccc}
  1 & 0 & 0 & 0 \\
  0 & 0 & 1 & 0 \\
  0 & 1 & 0 & 0 \\
  0 & 0 & 0 & 1 \\
\end{array}%
\right] =
$$
\begin{equation}\label{eq-sd-I-imm1to2q}
= \left[
\begin{array} {cccc}
a&0&b&0\\
0&a&0&b\\
c&0&d&0\\
0&c&0&d
\end{array}
\right] = U \otimes I;
\end{equation}
\end{itemize}
where $I$, here and in the following, denotes the identity matrix.

Observe that the analytical process in Eq. (\ref{eq-sd-I-imm1to2q}) can be immediately obtained from the corresponding diagrams of states, as shown in Figure \ref{sd-I-diastimm1to2q}.

As happens in the representations of generalized $\mbox{\sc cnot}$ gates, notice that the lines of the states to which the unitary matrix is applied do intersect the matrix representation, while the lines of the states to which the unitary matrix is not applied are overlapping and not intersecting.
The appropriate application of $\mbox{\sc swap}$ gates allows one to visualize the same computation, without intersections or overlaps of matrix or state lines, as shown in Figure \ref{sd-I-diastimm1to2q}.

\subsubsection{Immersions of \mbox{\sc not} gates on the space of two qubits}

We illustrate the simple example of immersions of $\mbox{\sc not}$ gates on the space of two qubits. The $\mbox{\sc not}$ gate can be applied to the least significant qubit and to the most significant qubit, respectively:
\begin{equation}
    \mbox{\sc not}_{\;LSB} = \left[\begin{array}{cccc}0&1&0&0\\1&0&0&0\\0&0&0&1\\0&0&1&0
\end{array}\right]; \mh \mbox{\sc not}_{\;MSB} = \left[\begin{array}
{cccc}0&0&1&0\\0&0&0&1\\1&0&0&0\\0&1&0&0\end{array}\right].
\end{equation}

\begin{figure}[!htb]
\begin{center}
\includegraphics[width=9cm]{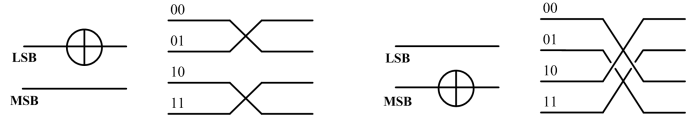}
\end{center}
  \caption{Two-qubit quantum operations and the corresponding elementary diagrams of states: immersions of $\mbox{\sc not}$ gates on the space of two qubits. From left to right, the $\mbox{\sc not}$ gate is applied to the least significant qubit and to the most significant qubit.} \label{sd-I-diastimmnot}
\end{figure}

The quantum circuits and corresponding diagrams of states are shown in Figure \ref{sd-I-diastimmnot}.

\subsection{Controlled gates}
\begin{figure}[htb]
\begin{center}
\includegraphics[width=9cm]{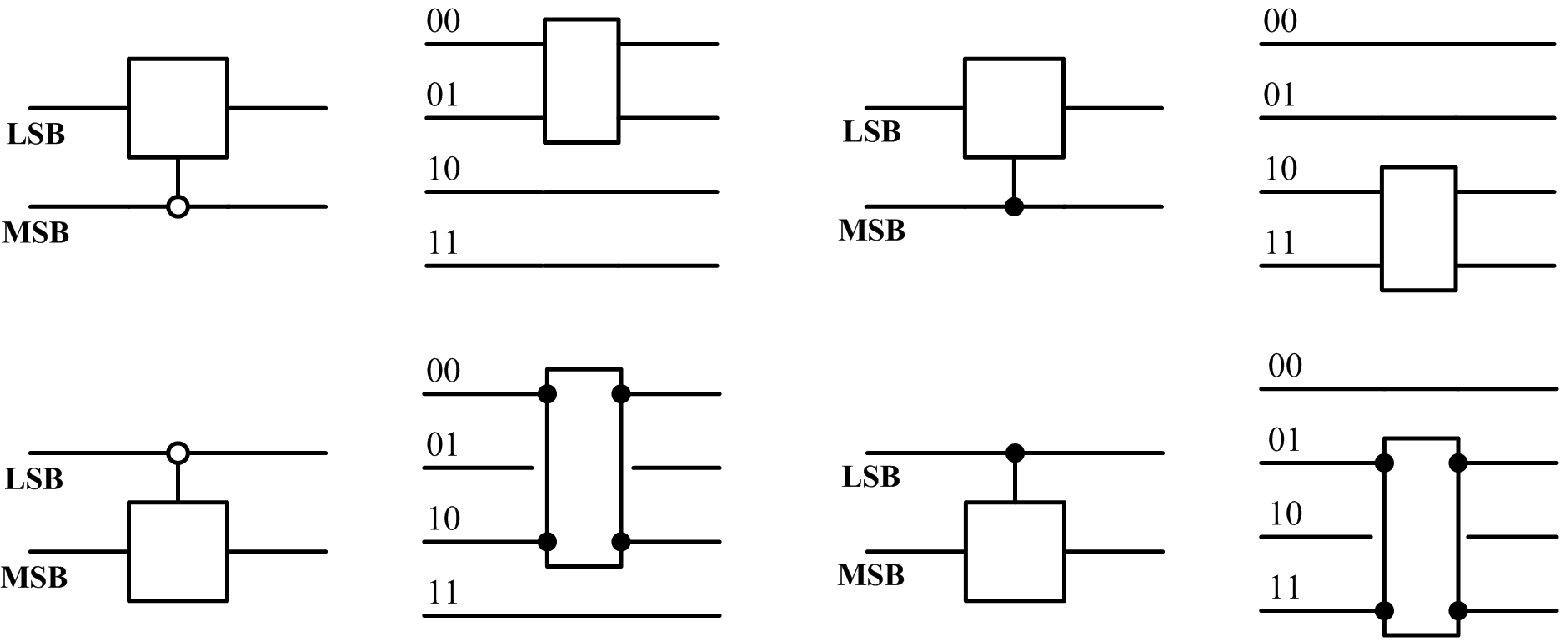}
\end{center}
  \caption{Two-qubit quantum operations and the corresponding elementary diagrams of states: controlled gates. As in the previous representations by means of diagrams of states, the lines of the states to which the unitary matrix is applied do intersect the matrix representation. Moreover, here and in the following figures, whenever useful, the points of junction are additionally marked.} \label{sd-I-diastcontr2q0}
\end{figure}

We consider the four possible controlled gates with the control from a single qubit and we show their representations by means of quantum circuits and diagrams of states in Figure \ref{sd-I-diastcontr2q0}.

\begin{figure}[htb]
\begin{center}
\includegraphics[width=8.2cm]{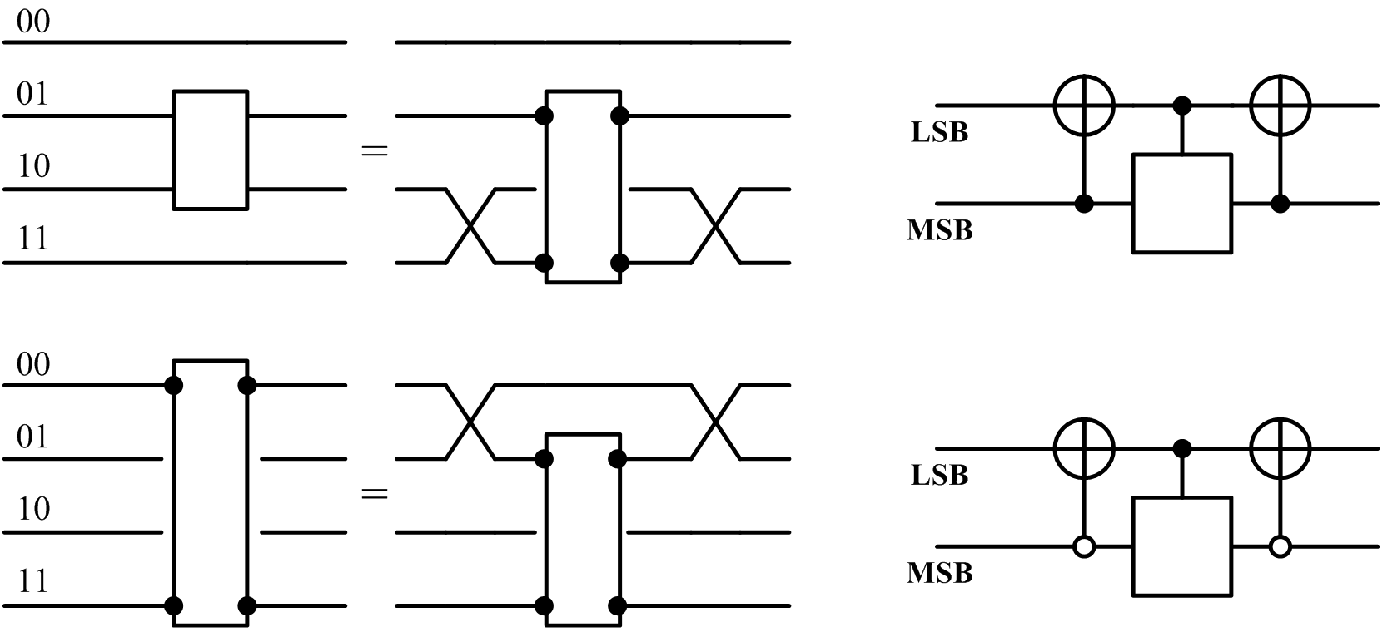}
\end{center}
  \caption{Two-qubit quantum operations and the corresponding elementary diagrams of states: two further examples of controlled gates. The diagrams of states (left) can be easily related to the fourth diagram in Figure \ref{sd-I-diastcontr2q0}, by the appropriate application of generalized $\mbox{\sc cnot}$ gates, as shown in the corresponding quantum circuits (right).}
\label{sd-I-diastcontr2q1}
\end{figure}

Figure \ref{sd-I-diastcontr2q1} completes the set of all possible controlled gates. These last representations can be easily related to the fourth diagram in Figure \ref{sd-I-diastcontr2q0}, by the appropriate application of generalized $\mbox{\sc cnot}$ gates. The diagrams of states allow one to immediately define the necessary combinations of controlled gates and generalized $\mbox{\sc cnot}$ gates, requiring no further analytical study.

\section{Elementary Three-qubit Operations}\label{sez-sd-I-3q}
Three qubits constitute an eight-state quantum system. In the following, we illustrate the immersions of two-qubit gates on the space of three qubits and the corresponding diagrams of states. Useful specific examples, like immersions of the $\mbox{\sc swap}$ and the generalized $\mbox{\sc cnot}$ gates on the space of three qubits, will be analyzed in future works \cite{FeStsdII}.

Immersions can be easily generalized to design diagrams of states illustrating elementary operations in systems composed of a higher number of qubits.

The synthesis of general states of a three-qubit system will be presented in section  \ref{sez-sd-I-sint2q3q}.

\subsection{Immersions of two-qubit gates on the space of three qubits}
In this section we illustrate the immersions of two-qubit gates on the space of three qubits. There are three possible cases, since the two-qubit gate can be applied to any combination of two out of three qubits.

Consider a general unitary matrix acting on two qubits:
$$U=\left[\begin{array}{cccc}a&b&c&d\\e&f&g&h\\i&j&k&l\\
m&n&o&p\end{array}\right].$$

\begin{figure}[htb]
\begin{center}
\includegraphics[width=10.2cm]{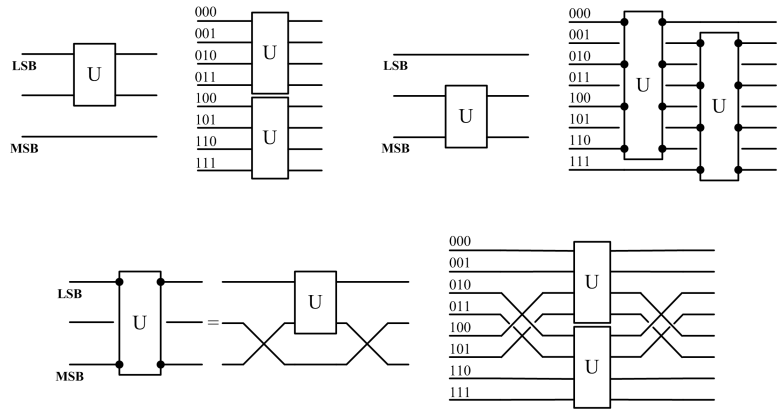}
\end{center}
  \caption{Three-qubit quantum operations and the corresponding elementary diagrams of states: immersions of two-qubit gates on the space of three qubits. From left to right, from top to bottom, numbering the qubits from the least significant to the most significant one, the unitary two-qubit matrix is applied to the qubits 0 and 1, 1 and 2, 0 and 2, respectively.}\label{sd-I-diastimm2to3q}
\end{figure}

Numbering the qubits from 0 to 2, from the least significant to the most significant one,
we obtain all possible immersions of the unitary matrix $U$ on the space of three qubits, as illustrated in Figure \ref{sd-I-diastimm2to3q}:

\begin{itemize}

\item[(i)] matrix $U$ applied to the qubits 0 and 1,
$$U_{01} = \left[
\begin{array}{cccccccc}a&b&c&d&0&0&0&0\\e&f&g&h&0&0&0&0\\
i&j&k&l&0&0&0&0\\m&n&o&p&0&0&0&0\\0&0&0&0&a&b&c&d\\0&0&0&0&e&f&g&h\\
0&0&0&0&i&j&k&l\\0&0&0&0&m&n&o&p\end{array}\right] = I \otimes U;$$

\item[(ii)] matrix $U$ applied to the qubits 1 and 2,
$$U_{12} = \left[
\begin{array}{cccccccc}a&0&b&0&c&0&d&0\\0&a&0&b&0&c&0&d\\
e&0&f&0&g&0&h&0\\0&e&0&f&0&g&0&h\\i&0&j&0&k&0&l&0\\0&i&0&j&0&k&0&l\\
m&0&n&0&o&0&p&0\\0&m&0&n&0&o&0&p\end{array}\right] = U \otimes I;$$

\item[(iii)] matrix $U$ applied to the qubits 0 and 2, which can be immediately obtained from the previous case (i), by the appropriate application of two $\mbox{\sc swap}$ gates,
$$U_{02} = \left[
\begin{array}{cccccccc}a&b&0&0&c&d&0&0\\e&f&0&0&g&h&0&0\\
0&0&a&b&0&0&c&d\\0&0&e&f&0&0&g&h\\i&j&0&0&k&l&0&0\\m&n&0&0&o&p&0&0\\
0&0&i&j&0&0&k&l\\0&0&m&n&0&0&o&p\end{array}\right].$$

\end{itemize}

The diagrams in Figure \ref{sd-I-diastimm2to3q} show a useful application of the $\mbox{\sc swap}$ gate to the class of diagrams that illustrates immersions of quantum gates on spaces constituted by a higher number of qubits. The appropriate applications of $\mbox{\sc swap}$ gates allows one to shift the qubits, on which the gate is applied, to adjacent and less significant positions. This simplifies the graphic representations, \textit{i.e.} visualizes equivalent computations, reducing to the minimum the number of state lines intersecting the matrix representation, as shown in the bottom diagram in Figure \ref{sd-I-diastimm2to3q}.

Finally, as happens for immersions in the space of two qubits, the diagrams of states clearly show how the distribution of information is determined by the \qq weight'' of the qubit to which the operation is applied, when tensor products of matrices are performed.

\section{Controlled Unitary Matrices and Applications}\label{sez-sd-I-sintucont}
In this section we illustrate some representations and synthesis procedures for controlled unitary matrices, with control from a single qubit or a couple of qubits. These gates provide a useful synthesis of general unitary matrices acting on a two-qubit system.

Meaningful examples of controlled gates are offered by the controlled-phase gate, with one or two control qubits, and by the $\mbox{\sc c}^2\mbox{\sc not}$, also known as the Toffoli gate \cite{Tof80,FrTo82}.

In the following, we present some possible syntheses by means of elementary gates, both to show that the illustrated gates can effectively be realized and to offer well defined procedures to synthesize them.

The choice of the most convenient elementary gates to synthesize complex gates strictly depends on the physical implementation of the quantum computer. Thus, given the present state-of-the-art of quantum computers physical implementations, it is useful to consider several possible choices of quantum gate synthesis.

\subsection{Controlled unitary matrices with control from a single qubit}
\begin{figure}[htb]
\begin{center}
\includegraphics[width=9cm]{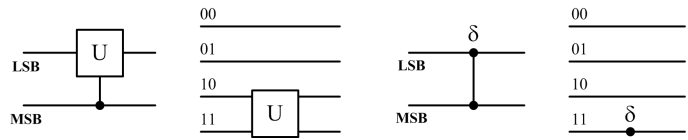}
\end{center}
  \caption{Controlled gates and their representation by diagrams of states: a general controlled unitary matrix (left) and the controlled phase-shift gate (right) with control from a single qubit.} \label{sd-I-diastcUph}
\end{figure}

A general controlled unitary matrix with control from a single qubit, denoted as $\mbox{\sc cU}$, is obtained by applying a single-qubit unitary matrix $U$ to the least significant qubit, only when the most significant qubit is in the state $\ket1$. Thus, a general controlled unitary matrix can be represented by the quantum circuit and diagram of states in Figure \ref{sd-I-diastcUph} (on the left).

Of course, these representations can be immediately generalized to obtain controlled gates acting on each one of the two (the least significant or the most significant) qubits and for each one of the two (\qq 0'' or \qq 1'') possible control values.

\subsubsection{Controlled phase-shift gates}
The controlled phase-shift gate is defined by:
\begin{equation}
    \mbox{\sc c}\delta = \left[\begin{array}{cccc}1&0&0&0\\0&1&0&0\\0&0&1&0\\0&0&0&e^{i\delta}
\end{array}\right]
\end{equation}
and it applies to the least significant qubit a phase-shift gate, previously illustrated, only when the most significant qubit is in the state $\ket1$. In other words, the phase shift $\delta$ is applied only to the component $\ket{11}$ of the overall state. The quantum circuit and diagram of states of the controlled phase-shift gate are illustrated in Figure \ref{sd-I-diastcUph} (on the right).

\begin{figure}[htb]
\begin{center}
\includegraphics[width=9.8cm]{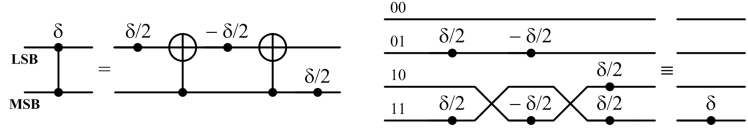}
\end{center}
  \caption{Controlled gates and their representation by diagrams of states: a synthesis of the controlled phase-shift gate with control from a single qubit.}
\label{sd-I-diastsintcph}
\end{figure}

Observe that the controlled phase-shift gate can be synthesized by the quantum circuit in Figure \ref{sd-I-diastsintcph}, by the appropriate application of two $\mbox{\sc cnot}$ gates and three phase-shift gates.

The diagram in Figure \ref{sd-I-diastsintcph} is the first example proposed in this tutorial of effective computations by means of the graphic representation of states: Without needing any further analytical study, interferences of quantum information can be immediately derived by following the action of the gates on the state lines.
Precisely, the phase-shift gates cause, on the one hand, destructive interferences on the states $\{\ket{01}, \ket{10}\}$ and, on the other hand, constructive interference on the state $\ket{11}$. Information flows, from left to right, along each state line and the effects of the phase-shift gates add up algebraically. Thus this diagram allows one to immediately compute the output state, instead of requiring five matrix multiplications, involving also three immersions of single-qubit gates.

\subsection{Synthesis of controlled unitary gates with control from a single qubit}
A general controlled special unitary matrix with control from a single qubit can be synthesized by the appropriate application of two $\mbox{\sc cnot}$ gates and three rotation gates, as it is well known in the literature (see, \textit{e.g.}, \cite{qcbook}, Ch. 3, pages 119-120).

Indeed, for any general special unitary matrix, there exist three unitary matrices $A, B, C$ such that:
\begin{equation}\label{eq-sd-I-cU}
    A  \; B \; C = I, \mh A  \; \mbox{\sc not}  \; B  \; \mbox{\sc not} \; C = SU.
\end{equation}

\begin{figure}[htb]
\begin{center}
\includegraphics[width=10.2cm]{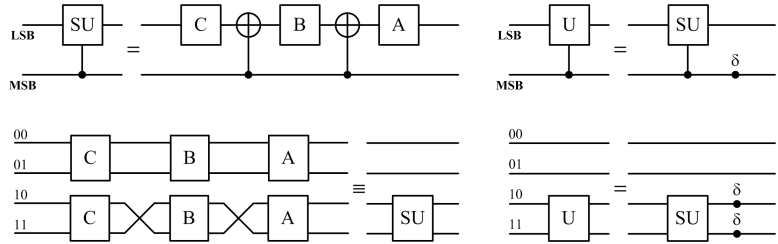}
\end{center}
  \caption{Controlled gates and their representation by diagrams of states: a synthesis of general controlled special unitary matrices (left) and of general controlled unitary matrices (right) with control from a single qubit.}
\label{sd-I-diastcU}
\end{figure}

Figure \ref{sd-I-diastcU} illustrates the quantum circuit and diagram of states for the synthesis of a controlled special unitary matrix with control on a single qubit, by means of the unitary matrices $A, B, C$, which satisfy the algebraic relations expressed in equation (\ref{eq-sd-I-cU}). The diagram of states, more clearly than the corresponding quantum circuit, shows the role of the algebraic relations imposed by Eq. (\ref{eq-sd-I-cU}).

Finally, to obtain a general controlled unitary matrix, it is sufficient to apply in sequence a controlled special unitary matrix and a phase-shift gate acting on the most significant qubit. This is equivalent to applying to the quantum circuit previously illustrated the gate $V(\delta)$, defined as follows:
\begin{equation}
    V(\delta) = \delta \otimes I = \left[%
\begin{array}{cccc}
  1 & 0 & 0 & 0 \\
  0 & 1 & 0 & 0 \\
  0 & 0 & e^{i \delta} & 0 \\
  0 & 0 & 0 & e^{i \delta} \\
\end{array}%
\right].
\end{equation}

\subsection{Synthesis of general two-qubit unitary matrices}
Several methods to synthesize general unitary matrices in a two-qubit system can be found in the literature. Among all possibilities, the synthesis by means of controlled unitary matrices and a particular matrix, generally denoted as $D_{0}$, can be very useful in several applications (see, \textit{e.g.}, \cite{qcbook}, Ch. 3, pages 124-126, and \cite{tucci}).

\begin{figure}[htb]
\begin{center}
\includegraphics[width=7.2cm]{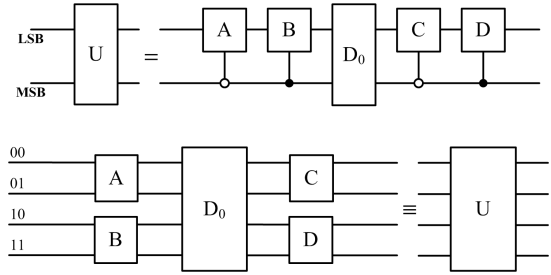}
\end{center}
  \caption{Controlled gates and their representation by diagrams of states: a synthesis of general unitary matrices in a two-qubit system. This synthesis is characterized by the sharp separation between controlled matrices (denoted by $A,
B, C, D$) and the particular matrix $D_{0}$.}
\label{sd-I-diastsintU2q}
\end{figure}

\begin{figure}[htb]
\begin{center}
\includegraphics[width=7.6cm]{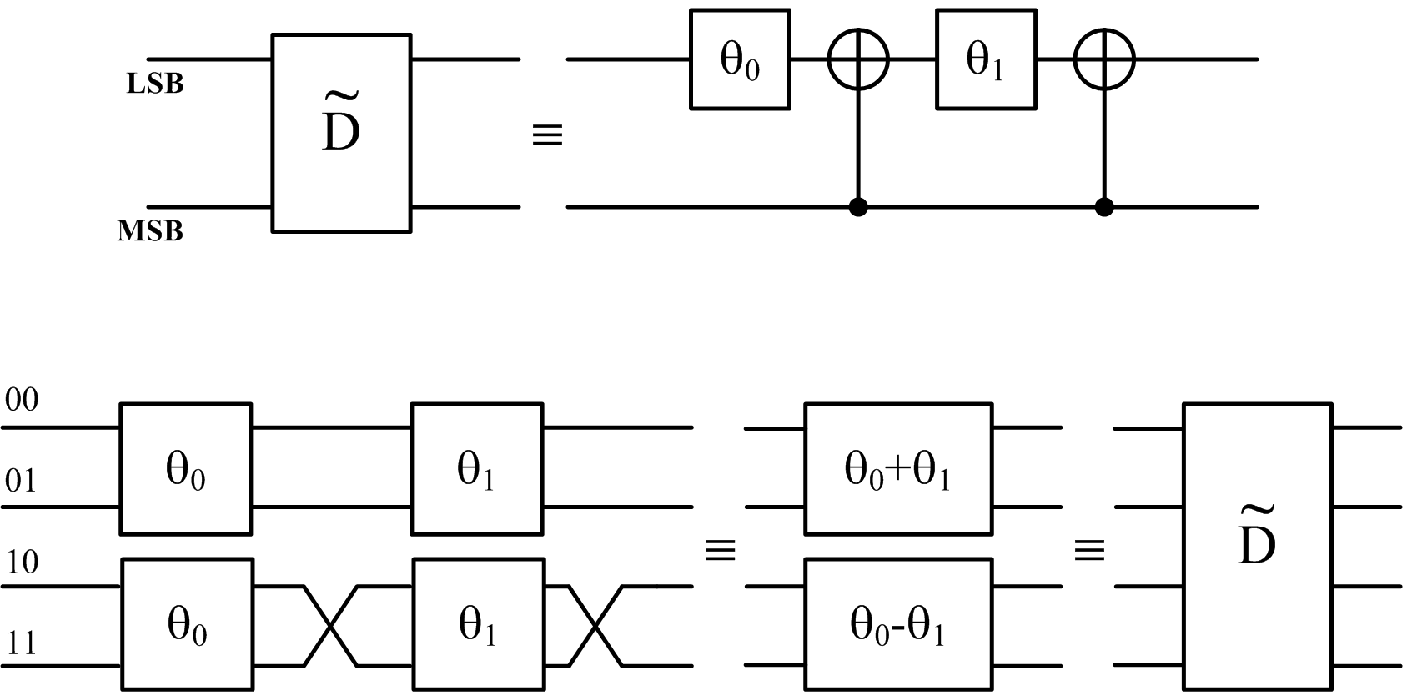}
\end{center}
  \caption{Controlled gates and their representation by diagrams of states: synthesis of the particular matrix $\tilde{D}$. The two upper state lines in the diagram of states, corresponding to the couple of states $\{00, 01\}$, are equivalent to a rotation by an angle $(\theta_0 + \theta_1)$, while the two lower state lines, corresponding to the couple of states $\{10, 11\}$, are equivalent to a rotation by an angle $(\theta_0 - \theta_1)$.} \label{sd-I-diastsintDtil}
\end{figure}

\begin{figure}[htb]
\begin{center}
\includegraphics[width=8.8cm]{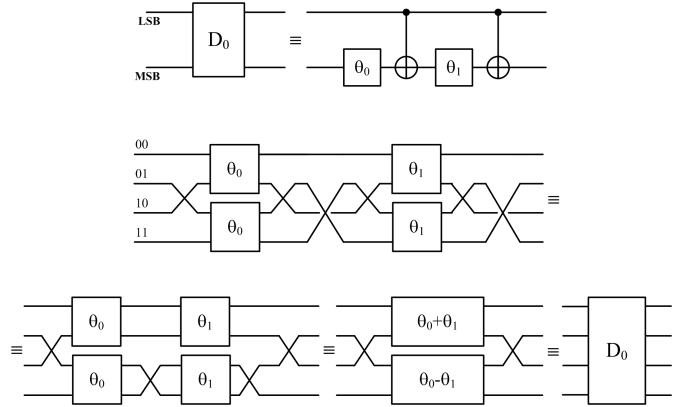}
\end{center}
  \caption{Controlled gates and their representation by diagrams of states: synthesis of the particular matrix $D_{0}$ for the decomposition of general unitary matrices in a two-qubit system, as shown in Figure \ref{sd-I-diastsintU2q}.}
\label{sd-I-diastsintDo}
\end{figure}

This synthesis can be obtained in three main steps:
\begin{itemize}
  \item [(i)] the initial two-qubit matrix is decomposed into four controlled unitary matrices and a two-qubit unitary matrix, $D_{0}$, of easier implementation (see Figure \ref{sd-I-diastsintU2q}). Notice that the dimension of each controlled unitary matrix is half the dimension of the initial matrix, \textit{i.e.} the four controlled gates are single-qubit unitary matrices;
  \item [(ii)] to synthesize the matrix $D_{0}$, we first synthesize an auxiliary matrix, denoted by $\tilde{D}$ (see Figure \ref{sd-I-diastsintDtil});
  \item [(iii)] finally, the matrix $D_0$ can be obtained by means of two $\mbox{\sc swap}$ gates and the matrix $\tilde{D}$, previously synthesized (see Figure \ref{sd-I-diastsintDo}).
\end{itemize}

The decomposition in step (i) is illustrated in Figure \ref{sd-I-diastsintU2q}. This procedure can be easily generalized for unitary matrices with dimension $2^n$, where the integer $n$ is the number of qubits that compose the system. The initial matrix with dimension $2^n$ is decomposed into four controlled unitary matrices and a unitary matrix of easier implementation. The dimension of each controlled unitary matrix is half the dimension of the initial matrix, that is $2^{n-1}$. By subsequent iterations, the initial matrix is finally decomposed into controlled unitary matrices acting on a two-qubit system and particular matrices of easier implementation.

The quantum circuit and corresponding diagram of states of step (ii) are illustrated in Figure \ref{sd-I-diastsintDtil}. According to the notation:
\begin{equation}\label{eq-sd-I-cossen01}
    C_0 = \cos \theta_0, \mh S_0 = \sin \theta_0, \mh C_1 =
    \cos \theta_1, \mh S_1 = \sin \theta_1,
\end{equation}
the complete diagram of states (in the middle) can be simplified (on the right) as expressed by the following equation, involving only the two lower state lines in the diagram, which correspond to the couple of states $\{10, 11\}$:
$$
\left[\begin{array}{cc} 0&1\\1&0
\end{array}
\right] \left[\begin{array}{cc} C_1& - S_1\\S_1&C_1
\end{array}
\right] \left[\begin{array}{cc} 0&1\\1&0
\end{array}
\right] \left[\begin{array}{cc} C_0& - S_0\\S_0&C_0
\end{array}
\right]=
$$
\begin{equation}\label{eq-sd-I-Dosimpl}
= \left[\begin{array}{cc}
\cos(\theta_0 - \theta_1)& - \sin(\theta_0 - \theta_1)\\
\sin(\theta_0 - \theta_1)& \cos(\theta_0 - \theta_1)
\end{array}
\right].
\end{equation}

Thus the sequence of gates acting on the couple of states $\{10, 11\}$ is equivalent to a rotation of an angle $(\theta_0 - \theta_1)$. On the other hand, by observing the complete diagram in Figure \ref{sd-I-diastsintDtil}, we can immediately guess that the overall transformation affecting the two upper state lines, corresponding to the couple of states $\{00, 01\}$, is equivalent to a rotation by an angle $(\theta_0 + \theta_1)$.

Thus, denoting:
$$C_a = \cos(\theta_0 + \theta_1), \mh S_a = \sin(\theta_0 +
\theta_1),$$
\begin{equation}\label{eq-sd-I-cossenab} C_b =
\cos(\theta_0 - \theta_1), \mh S_b = \sin(\theta_0 - \theta_1),
\end{equation}
we obtain the overall matrix $\tilde{D}$:
\begin{equation}
\tilde{D} = \left[\begin{array}{cccc} C_a& -
S_a&0&0\\S_a&C_a&0&0\\0&0&C_b& - S_b\\0&0&S_b&C_b
\end{array}\right].
\end{equation}

Finally, the matrix $D_0$ can be synthesized by means of the quantum circuit and corresponding diagram of states illustrated in Figure \ref{sd-I-diastsintDo}. The latter can be simplified, as previously shown in Figure \ref{sd-I-diastsintDtil}.

By means of two $\mbox{\sc swap}$ gates and the matrix $\tilde{D}$, the matrix $D_0$ can be obtained as follows:
$$
D_0 = \left[\begin{array}{cccc} 1&0&0&0\\0&0&1&0\\0&1&0&0\\0&0&0&1
\end{array}\right]
\left[\begin{array}{cccc} C_a& - S_a&0&0\\S_a&C_a&0&0\\0&0&C_b& -
S_b\\0&0&S_b&C_b
\end{array}\right]
\left[\begin{array}{cccc} 1&0&0&0\\0&0&1&0\\0&1&0&0\\0&0&0&1
\end{array}\right]=
$$
\begin{equation}
= \left[\begin{array}{cccc} C_a&0& - S_a&0\\0&C_b&0& -
S_b\\S_a&0&C_a&0\\0&S_b&0&C_b
\end{array}\right].
\end{equation}

\subsection{Controlled unitary matrices with control from a couple of qubits}
\begin{figure}[htb]
\begin{center}
\includegraphics[width=10.4cm]{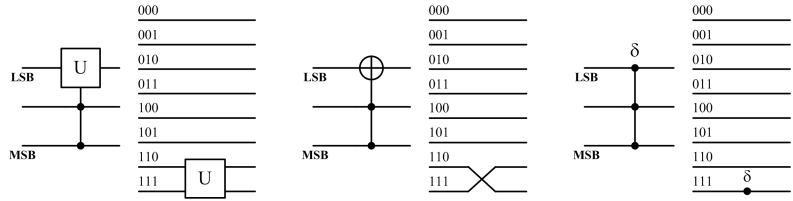}
\end{center}
  \caption{Controlled gates and their representation by diagrams of states: a general controlled unitary matrix with control from the two most significant qubits (left), the Toffoli gate (center) and the controlled phase-shift gate with control from the two most significant qubits (right).} \label{sd-I-diastc2UTph}
\end{figure}

A general controlled unitary matrix with control from the two most significant qubits, denoted as $\mbox{\sc c}^2\mbox{\sc U}$, is obtained by applying a single-qubit unitary matrix $U$ to the least significant qubit, only when both the two most significant qubits are in the state $\ket1$. Thus, a general controlled unitary matrix with control from the two most significant qubits can be represented by the quantum circuit and diagram of states shown in Figure \ref{sd-I-diastc2UTph} (on the left).

By means of the appropriate application of $\mbox{\sc not}$ gates on the control qubits, these representations can be immediately generalized to obtain controlled gates active on any possible combination of values of the control qubits (the value of each qubit is chosen, as usual, to be \qq 0'' or \qq 1'').

\subsubsection{The Toffoli gate}
The Toffoli gate, denoted also as $\mbox{\sc c}^2\mbox{\sc not}$, applies a $\mbox{\sc not}$ gate to the least significant qubit, only when both the two most significant qubits are in the state $\ket1$. Thus, the Toffoli gate can be represented by the quantum circuit and diagram of states shown in Figure \ref{sd-I-diastc2UTph} (center).

\subsubsection{Controlled phase-shift gates with control from a couple of qubits}
The controlled phase-shift gate with control from the two most significant qubits, acting on a three-qubit system, applies to the least significant qubit a phase-shift gate, previously illustrated,  only when both the two most significant qubits are in the state $\ket1$. In other words, the phase shift $\delta$ is applied only to the component $\ket{111}$ of the overall state.

\begin{figure}[htb]
\begin{center}
\includegraphics[width=9.6cm]{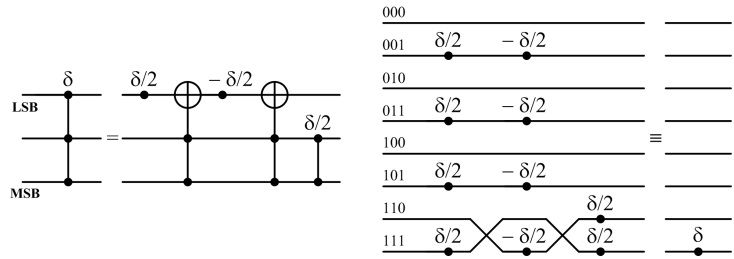}
\end{center}
  \caption{Controlled gates and their representation by diagrams of states: a synthesis of the controlled phase-shift gate with control from the two most significant qubits.} \label{sd-I-diastcph3q}
\end{figure}

The quantum circuit and diagram of states of the controlled phase-shift gate with control from the two most significant qubits are illustrated in Figure \ref{sd-I-diastc2UTph} (on the right). Observe that this controlled phase-shift gate  can be synthesized by the appropriate application of two $\mbox{\sc c}^2\mbox{\sc not}$ gates, two phase-shift gates and a controlled phase-shift gate acting on a two-qubit system, as shown in the quantum circuit of Figure \ref{sd-I-diastcph3q}.

Again, we observe a meaningful example of effective computation by means of the graphic representation of states: Without needing any further analytical study, interference of quantum information can be immediately derived by following the action of the gates on the state lines.
Precisely, the phase-shift gates cause, on the one hand, destructive interferences on the states $\{001, 011, 101, 110\}$ and, on the other hand, constructive interference on the state $111$. Information flows, from left to right, along each state line and the effects of the phase-shift gates add up algebraically.

\subsection{Synthesis of controlled unitary gates with control from a couple of qubits}
A general controlled unitary matrix with control from a couple of qubits, acting on a three-qubit system, can be synthesized by the appropriate application of $\mbox{\sc cnot}$ gates and controlled gates with control from a single qubit, as it is well known in the literature (see, \textit{e.g.}, \cite{qcbook}, Ch. 3, pages 120-121).

\begin{figure}[htb]
\begin{center}
\includegraphics[width=6.8cm]{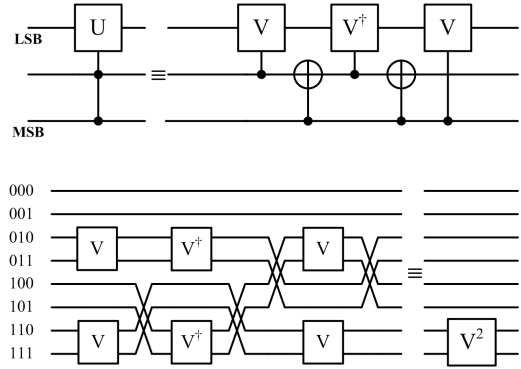}
\end{center}
  \caption{Controlled gates and their representation by diagrams of states: a synthesis of general controlled unitary matrices with control from the two most significant qubits, acting on a three-qubit system.} \label{sd-I-diastc2U}
\end{figure}

Figure \ref{sd-I-diastc2U} illustrates the quantum circuit and diagram of states for the synthesis of a controlled unitary matrix with control from the two most significant qubits, denoted as $\mbox{\sc c}^2\mbox{\sc U}$. Two $\mbox{\sc cnot}$ gates and three controlled gates are necessary: Among the controlled gates, we have two $\mbox{\sc cV}$ and one $\mbox{\sc cV}^\dagger$, where the unitary matrix $V$ is such that $V^2 = U$.

Without needing any further analytical study, the diagram of states highlights the role of the algebraic relation $V^2 = U$ more straightforwardly than the corresponding quantum circuit. The destructive and constructive interferences of quantum information can be easily visualized by following the action of the gates applied in sequence.

\section{Diagonal Unitary Matrices and Applications}\label{sez-sd-I-sintudiag}
In this section we illustrate some synthesis procedures for diagonal unitary matrices in two-qubit and three-qubit systems. This method can be easily generalized to synthesize diagonal unitary matrices acting on an $m$-qubit system.
These matrices can be synthesized by means of the appropriate application of phase-shift gates and controlled phase-shift gates, previously illustrated. The diagrams of states allow one to determine the matrix that transforms the parameters of the synthesis circuit into the parameters defining the general diagonal unitary matrix.

Finally, we illustrate a possible synthesis of general two-qubit and three-qubit states; the procedure can be easily generalized to synthesize quantum states of systems composed of a higher number of qubits.

\subsection{Diagonal two-qubit unitary matrices}\label{sez-sd-I-diag2q}
\begin{figure}[htb]
\begin{center}
\includegraphics[width=6.4cm]{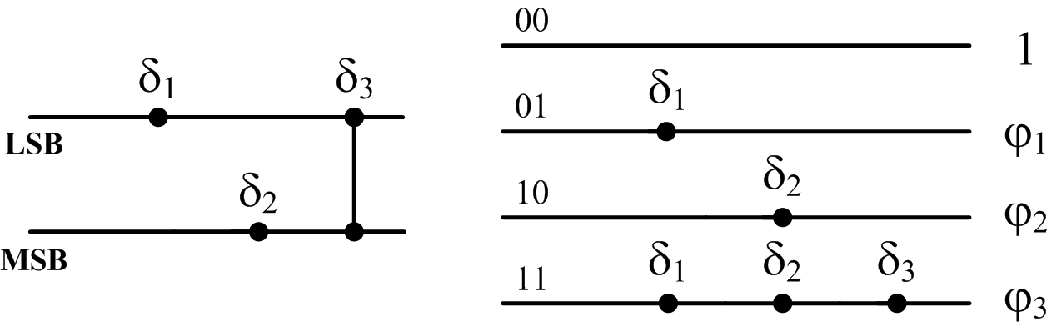}
\end{center}
  \caption{Diagonal unitary matrices and their representation by diagrams of states: synthesis of diagonal unitary matrices in a two-qubit system. The phase-shift gates cause constructive interference in the state $\ket{11}$, thus determining the parameter $\varphi_3$.}
\label{sd-I-diastdiag2q}
\end{figure}

We illustrate a possible synthesis of a diagonal unitary matrix in a two-qubit system in Figure \ref{sd-I-diastdiag2q}, which shows the corresponding quantum circuit and diagram of states.

Diagonal unitary matrices are assumed to be defined with an arbitrary common phase factor. Neglecting such arbitrary common phase factor, a general diagonal unitary matrix in a two-qubit system is defined by three free parameters:
\begin{equation}
    D_2=\left[\begin{array}{cccc}1&0&0&0\\0&e^{i\phi_1}&0&0\\0&0&e^{i\phi_2}&0\\
0&0&0&e^{i\phi_3}\end{array}\right].
\end{equation}

The following expressions can be immediately derived from the diagram of states, without needing any further analytical study:
\begin{equation}
\delta_1 = \varphi_1; \mh \delta_2 = \varphi_2; \mh
\delta_1 + \delta_2 + \delta_3 = \varphi_3.
\end{equation}

\subsection{Diagonal three-qubit unitary matrices}\label{sez-sd-I-diag3q}
\begin{figure}[htb]
\begin{center}
\includegraphics[width=9.6cm]{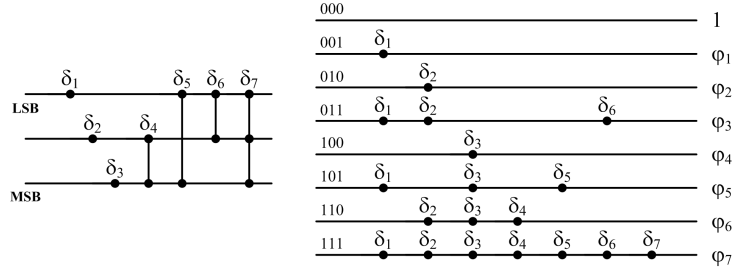}
\end{center}
  \caption{Diagonal unitary matrices and their representation by diagrams of states: synthesis of diagonal unitary matrices for a three-qubit system. The phase-shift gates cause constructive interference, thus determining the parameters $\{\varphi_i: i = 1,...,7$\}.}
\label{sd-I-diastdiag3q}
\end{figure}

We illustrate a possible synthesis of a diagonal unitary matrix for a three-qubit system, by means of phase-shift gates, two-qubit and three-qubit controlled phase-shift gates. Figure \ref{sd-I-diastdiag3q} shows the corresponding quantum circuit and diagram of states.

Neglecting an arbitrary common phase factor, a general diagonal unitary matrix in a three-qubit system can be defined by seven free parameters:
\begin{equation}
    D_3 = diag [1, e^{i\varphi_1}, e^{i\varphi_2}, e^{i\varphi_3},
e^{i\varphi_4}, e^{i\varphi_5}, e^{i\varphi_6}, e^{i\varphi_7}].
\end{equation}

The following expressions for the parameters $\{\delta_i, \varphi_i : i = 1,...,7$\} can be immediately derived from the diagram of states, without needing any further analytical study:
$$
\delta_1 = \varphi_1; \mh \delta_2 = \varphi_2; \mh
\delta_3 = \varphi_4;
$$
$$
\delta_1 + \delta_2 + \delta_6 = \varphi_3; \mh \delta_1 +
\delta_3 + \delta_5 = \varphi_5; \mh \delta_2 + \delta_3 +
\delta_4 = \varphi_6;
$$
\begin{equation}
\delta_1 + \delta_2 + \delta_3 + \delta_4 + \delta_5 + \delta_6 +
\delta_7 = \varphi_7.
\end{equation}

From the previous expressions, the inverse relations can be derived, obtaining:
$$
\delta_1 = \varphi_1; \mh \delta_2 = \varphi_2; \mh
\delta_3 = \varphi_4;
$$
$$
\delta_4 = \varphi_6 - \varphi_2 - \varphi_4; \mh \delta_5 =
\varphi_5 - \varphi_1 - \varphi_4; \mh \delta_6 = \varphi_3 -
\varphi_1 - \varphi_2;
$$
\begin{equation}
\delta_7 = \varphi_7 + \varphi_1 + \varphi_2 + \varphi_4 -
\varphi_3 - \varphi_5 - \varphi_6.
\end{equation}

\subsection{Synthesis of two-qubit and three-qubit states}\label{sez-sd-I-sint2q3q}
A general state of a two-qubit system, a three-qubit system or a system composed of a higher number of qubits can be generated by separately synthesizing the desired amplitude moduli and phases.

The synthesis of the phases of general two-qubit and three-qubit states can be performed by synthesizing general diagonal unitary matrices in two-qubit and three-qubit systems, as previously illustrated in Sections~\ref{sez-sd-I-diag2q} and
\ref{sez-sd-I-diag3q}, respectively.

Thus, in the following, we illustrate in detail how to synthesize the amplitude moduli of entangled two-qubit and three-qubit states. These procedures can be easily generalized to synthesize the amplitude moduli of entangled states in systems composed of a higher number of qubits.

\subsubsection*{Synthesis of the amplitudes of a two-qubit state}
\begin{figure}[htb]
\begin{center}
\includegraphics[width=7.6cm]{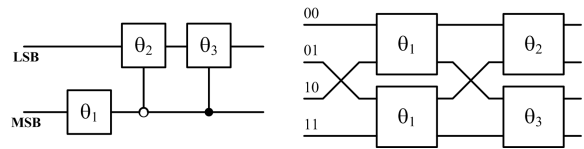}
\end{center}
  \caption{Quantum circuit and diagram of states for the synthesis of the amplitude moduli of a general entangled two-qubit state.} \label{sd-I-diastst2q}
\end{figure}

The synthesis of the amplitude moduli of a general entangled two-qubit state can be obtained by the quantum circuit and corresponding diagram of states illustrated in Figure \ref{sd-I-diastst2q}. The controlled gate $\theta_2$ acts on the states for which $\textsc{msb} = 0$, thus it is applied to the top state lines in the diagram. On the other hand, the controlled gate $\theta_3$ acts on the states for which $\textsc{msb} = 1$, thus it is applied to the bottom state lines in the diagram.

\subsubsection*{Synthesis of the amplitudes of a three-qubit state}
\begin{figure}[htb]
\begin{center}
\includegraphics[width=8cm]{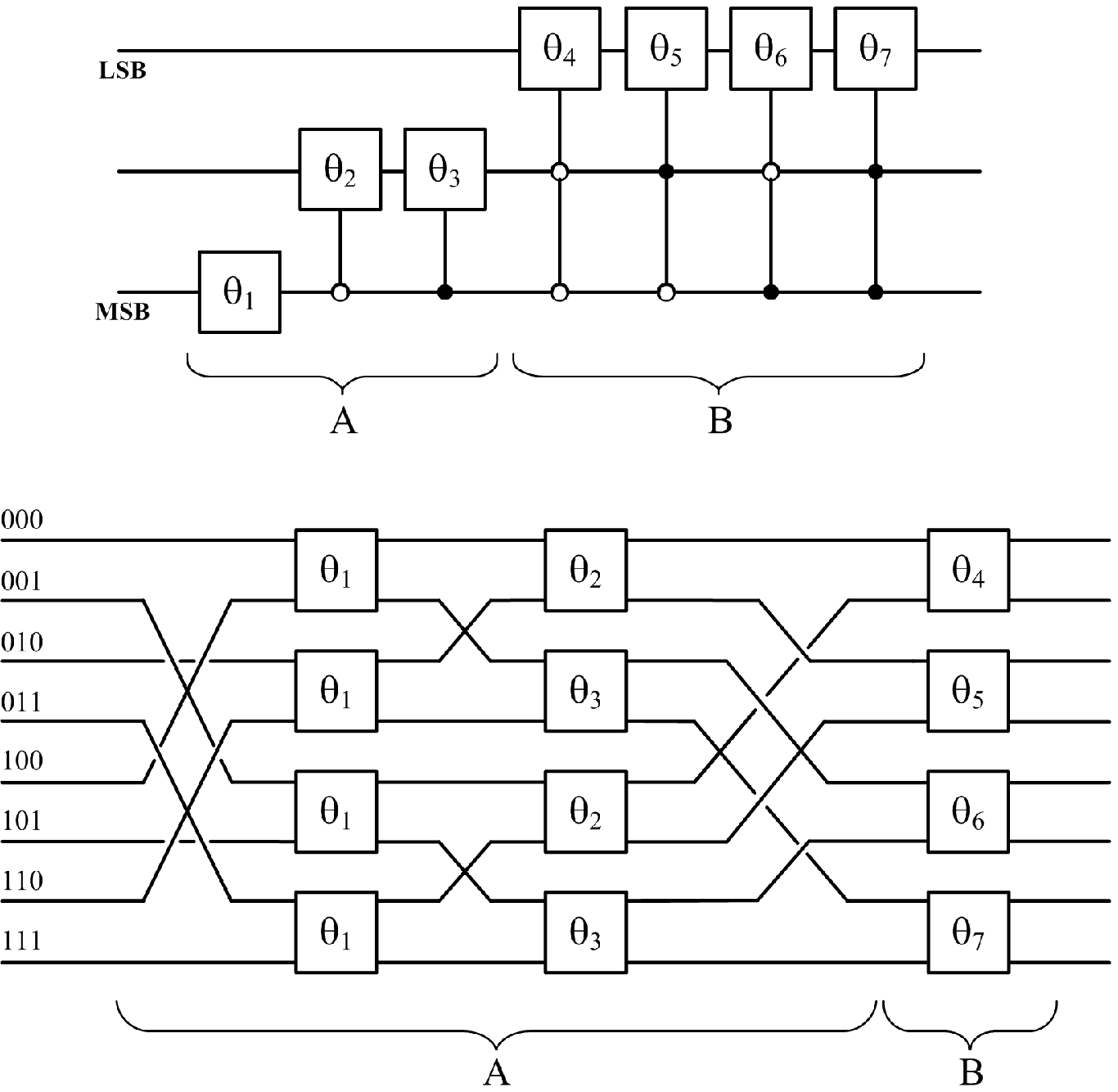}
\end{center}
  \caption{Quantum circuit and diagram of states for the synthesis of the amplitude moduli of a general entangled three-qubit state. The sequence of gates for the synthesis of the amplitude moduli of a general entangled two-qubit state, denoted by \qq A'' and illustrated in Figure \ref{sd-I-diastst2q}), is applied to the two most significant qubits; subsequently, all possible controlled gates with control from a couple of qubits are applied in sequence (denoted by \qq B'').} \label{sd-I-diastst3q}
\end{figure}

The synthesis of the amplitude moduli of a general entangled three-qubit state can be obtained by the quantum circuit and corresponding diagram of states illustrated in Figure \ref{sd-I-diastst3q}. The first part of the quantum circuit (denoted by \qq A'' in Figure \ref{sd-I-diastst3q}) applies to the two most significant qubits the sequence of gates for the synthesis of the amplitude moduli of a general entangled two-qubit state, illustrated in Figure \ref{sd-I-diastst2q}. Subsequently, all possible controlled gates with control from a couple of qubits are applied in sequence; their application is ordered in respect to the control value, from \qq 0'' to \qq 1'', from the least significant qubit to the most significant qubit (this sequence of gates is denoted by \qq B'' in Figure \ref{sd-I-diastst3q}).

In the complete diagram of states, the sequence of gates for the synthesis of the amplitude moduli of a general entangled two-qubit state, set into the space of three qubits, is also denoted by \qq A''. After this sequence of gates, the controlled gates with control from a couple of qubits are applied to couples of states corresponding to the four possible combinations of the control values (also denoted by \qq B''). As clearly shown by the diagram, the controlled gate $\theta_4$ acts on the states for which both the two \textsc{msb} are equal to \qq 0'', \textit{i.e.} the couple of states $\{000, 001\}$; the controlled gate $\theta_5$ acts on the states for which the two \textsc{msb} are equal to \qq 0'' and \qq 1'', respectively, \textit{i.e.} the couple of states $\{010, 011\}$; the controlled gate $\theta_6$ acts on the states for which the two \textsc{msb} are equal to \qq 1'' and \qq 0'', respectively, \textit{i.e.} the couple of states $\{100, 101\}$; the controlled gate $\theta_7$ acts on the states for which both the two \textsc{msb} are equal to \qq 1'', \textit{i.e.} the couple of states $\{110, 111\}$.

For the sake of simplicity, we omit the corresponding analytical expressions.

\section{Conclusions and Future Developments}\label{sez-sd-I-concl}
In the present tutorial we have introduced and described in detail the \textit{Diagrams of States}, a new (to the best of our knowledge) way to graphically represent and analyze how quantum information is elaborated during execution of quantum circuits.

In our opinion, the diagrams of states can serve as an alternative approach to analyze known quantum algorithms, as well as an auxiliary tool to conceive novel quantum computations. In fact, they can be used in addition to traditional tools such as analytical study and Feynman diagrams, as these representations are too synthetic to clearly visualize how quantum information flows during computations. On the contrary, the dimension of the graphic representation of states grows exponentially in respect to the dimension of the quantum system to be described, but this feature has proven to be a merit rather than a flaw, since it has allowed a clearer visualization of every detail of the quantum processes considered so far.

The method of diagrams of states has been previously applied to study and compare some models of quantum copying machines \cite{FeSt06}. In this tutorial we have thus offered a complete and detailed illustration of this representation, by means of a constructive procedure including several applications and useful examples of quantum computations. We have illustrated elementary operations performed in single-qubit, two-qubit and three-qubit systems, immersions of quantum gates on systems composed of a higher number of qubits, generation of general multi-qubit states and procedures to synthesize unitary, controlled and diagonal matrices.

The diagrams of states prove to be most useful whenever the quantum operations to be analyzed are described by very sparse matrices and known and widely used quantum algorithms actually involve operations that satisfy this requirement. In such cases, each non null entry of a matrix is associated to a line in the matrix diagram, along which the information flows and is elaborated, while null entries of the matrix correspond to absence of matrix lines in diagram. Hence, the resulting diagrams show clearly and immediately how the quantum information flow should be \qq read'' during the described computations.

Moreover, for any given quantum computation, the graphic representation of states offers both a complete description of each gate's action on the states (the complete diagram) and an overall description of the transformation from the input to the output state (the simplified diagram). In fact, in order to perform the analysis of the quantum processes the complete diagrams of states have been directly derived from the physical implementation of the quantum circuits associated to the processes. Then these diagrams have been easily rearranged into the corresponding simplified diagrams, which better visualized the overall effects of the computations.

Finally, by considering a sort of inverse process, the diagrams of states could help to conceive new quantum algorithms. In fact, one could schematically describe the desired manipulation of quantum information by means of intuitive diagrams. These simplified diagrams could then help to guess the equivalent complete diagrams from which the corresponding implementations by means of quantum circuits could be obtained effortlessly.

Several meaningful examples of such synthesis, as well as the analysis of more complex algorithms and processes will be provided in following papers, for whose comprehension the present tutorial offers the necessary didactic introduction. We will illustrate the quantum states representation by density matrices, partial measurement and partial trace operations, density matrix purification and evolution according to the Kraus representation, generation and measurement of entangled states. We will also apply the method to widely known, more complex and meaningful quantum algorithms, such as the Deutsch's problem, the quantum teleportation and dense coding protocols, a general single-qubit decoherence model, entanglement distillation and quantum error correcting codes.

\section*{Acknowledgments}
The authors wish to thank Giuliano Benenti, Samuel L. Braunstein, Ignazio Licata and Roberto Suardi for their kind contributions to the development and improvement of this tutorial.


\end{document}